\documentclass[%
 reprint,
superscriptaddress,
 amsmath,amssymb,
 aps,
 prl,
]{revtex4-2}

\usepackage{graphicx}
\usepackage{float}
\usepackage{graphicx}
\usepackage{dcolumn}
\usepackage{bm}
\usepackage{verbatim}
\usepackage[dvipsnames]{xcolor}
\usepackage{hyperref}
\usepackage[mathlines]{lineno}
\usepackage{makecell}
\usepackage{comment}
\usepackage{siunitx}
\begin{document}
\preprint{APS/123-QED}
\title{Quantum teleportation over a field-deployed hollow-core fibre network}
\author{Ri-Yao Song}
\thanks{These authors contributed equally to this work.}
\affiliation{Institute of Fundamental and Frontier Sciences, University of Electronic Science and Technology of China, Chengdu 611731, China}
\affiliation{Center for Quantum Internet, Tianfu Jiangxi Laboratory, Chengdu 641419, China}
\affiliation{Key Laboratory of Quantum Physics and Photonic Quantum Information, Ministry of Education, University of Electronic Science and Technology of China, Chengdu 611731, China}
\author{Ya-Zhou Zhao}
\thanks{These authors contributed equally to this work.}
\affiliation{Institute of Fundamental and Frontier Sciences, University of Electronic Science and Technology of China, Chengdu 611731, China}
\affiliation{Center for Quantum Internet, Tianfu Jiangxi Laboratory, Chengdu 641419, China}
\affiliation{Key Laboratory of Quantum Physics and Photonic Quantum Information, Ministry of Education, University of Electronic Science and Technology of China, Chengdu 611731, China}
\author{Yun-Ru Fan}
\email{yunrufan@uestc.edu.cn}
\affiliation{Institute of Fundamental and Frontier Sciences, University of Electronic Science and Technology of China, Chengdu 611731, China}
\affiliation{Center for Quantum Internet, Tianfu Jiangxi Laboratory, Chengdu 641419, China}
\affiliation{Key Laboratory of Quantum Physics and Photonic Quantum Information, Ministry of Education, University of Electronic Science and Technology of China, Chengdu 611731, China}
\author{Yang-Bin Ma}
\affiliation{Institute of Fundamental and Frontier Sciences, University of Electronic Science and Technology of China, Chengdu 611731, China}
\affiliation{Center for Quantum Internet, Tianfu Jiangxi Laboratory, Chengdu 641419, China}
\affiliation{Key Laboratory of Quantum Physics and Photonic Quantum Information, Ministry of Education, University of Electronic Science and Technology of China, Chengdu 611731, China}
\author{Yan-Yu Wei}
\affiliation{Institute of Fundamental and Frontier Sciences, University of Electronic Science and Technology of China, Chengdu 611731, China}
\author{Si Shen}
\affiliation{Southwest Institute of Technical Physics, Chengdu 610041, China}

\author{Hao Li}
\affiliation{\mbox{National Key Laboratory of Materials for Integrated Circuits, Shanghai Institute of Microsystem and Information Technology,} \mbox{Chinese Academy of Sciences, Shanghai 200050, China}}
\author{Li-Xing You}
\affiliation{\mbox{National Key Laboratory of Materials for Integrated Circuits, Shanghai Institute of Microsystem and Information Technology,} \mbox{Chinese Academy of Sciences, Shanghai 200050, China}}
\author{Kai Guo}
\email{guokai07203@hotmail.com}
\affiliation{Institute of Systems Engineering, AMS Beijing 100141, China}
\author{Guang-Can Guo}
\affiliation{Institute of Fundamental and Frontier Sciences, University of Electronic Science and Technology of China, Chengdu 611731, China}
\affiliation{Center for Quantum Internet, Tianfu Jiangxi Laboratory, Chengdu 641419, China}
\affiliation{Key Laboratory of Quantum Physics and Photonic Quantum Information, Ministry of Education, University of Electronic Science and Technology of China, Chengdu 611731, China}
\affiliation{CAS Center for Excellence in Quantum Information and Quantum Physics, University of Science and Technology of China, Hefei 230026, China}
\author{Qiang Zhou}
\email{zhouqiang@uestc.edu.cn}
\affiliation{Institute of Fundamental and Frontier Sciences, University of Electronic Science and Technology of China, Chengdu 611731, China}
\affiliation{Center for Quantum Internet, Tianfu Jiangxi Laboratory, Chengdu 641419, China}
\affiliation{Key Laboratory of Quantum Physics and Photonic Quantum Information, Ministry of Education, University of Electronic Science and Technology of China, Chengdu 611731, China}
\affiliation{CAS Center for Excellence in Quantum Information and Quantum Physics, University of Science and Technology of China, Hefei 230026, China}

\date{\today}
\begin{abstract}
When a photon and one member of an entangled photon pair are jointly projected onto a Bell-state measurement (BSM), the quantum state of the photon can be transferred to the distant partner of the pair without physically transmitting this information carrier. In real-world deployment, however, teleportation performance is fundamentally bottlenecked by quantum channel impairments, such as loss, noise, and fluctuations, which induce severe decoherence and degrade fidelity. This vulnerability is further exacerbated in scenarios with intense classical data traffic or background light. Realizing scalable quantum networks, therefore, hinges on developing advanced channel architectures capable of supporting both high-fidelity quantum operations and high-capacity classical communications within a shared infrastructure. Towards this end, hollow core fibre (HCF) offers a promising quantum channel resource by combining free-space-like weak light-matter interaction with the stability of fibre-based systems. Here, utilizing a field-deployed metropolitan HCF network spanning three spatially separated nodes in Chengdu, we achieve quantum teleportation with an intermediate BSM under co-propagating classical traffic. Crucially, the HCF links preserve the long-term indistinguishability of photonic qubits without active stabilization, and exhibit a Raman noise approximately three orders of magnitude lower than that of standard solid-core counterparts. This noise suppression enables robust quantum teleportation even alongside classical launch powers up to 160 mW. Our findings establish a classical-data-compatible framework for quantum networking over deployed fibre infrastructure and offer a wavelength-agnostic, plug-and-play, and free-running pathway toward the quantum internet.

\end{abstract} 
\maketitle
Quantum teleportation \cite{bennett1993teleporting} has been demonstrated \cite{bouwmeester1997experimental,pan1998experimental,jennewein2001experimental,jia2004experimental} over diverse quantum channels (see Ref. \cite{lu2022micius,hu2023progress} for recent reviews), including metropolitan solid core fibre (SCF) links \cite{marcikic2003long,ursin2004quantum,de2004long,bussieres2014quantum,sun2016quantum,valivarthi2016quantum,huo2018deterministic,zhao2022real,shen2023hertz}, terrestrial free-space links \cite{jin2010expQT,yin2012quantum,ma2012quantum}, and satellite free-space links \cite{ren2017ground,li2022quantum}. The question of which channel resource is well-suited for deployment-ready teleportation-based quantum networks remains unexplored yet. Since the protocol requires independently prepared photonic qubits to maintain stringent indistinguishability after propagating through distinct links to undergo a joint Bell-state measurement (BSM), any link-induced polarization fluctuation, timing drift, or environmental noise inevitably undermines the interference success probability and degrades the teleported state fidelity. Standard SCF links are inherently susceptible to ambient perturbations that compromise polarization and temporal stability \cite{bussieres2014quantum,sun2016quantum,valivarthi2016quantum,valivarthi2020teleportation,shen2023hertz,iuliano2024qubit,thomas2024quantum,fan2025quantum,pittaluga2025long,liu2025chip,lu2026device,liu2026long,stas2026entanglement}. Free-space links largely eliminate material interactions, yet remain vulnerable to atmospheric turbulence and background noise \cite{jin2010expQT,yin2012quantum,ma2012quantum}. Satellite links extend quantum connectivity to global distances, but their operation is constrained by transient transmission windows dictated by orbital dynamics and meteorological conditions \cite{ren2017ground,li2022quantum,lu2022micius,li2025microsatellite}. These constraints become even more critical in realistic communication infrastructures, where fragile photonic qubits must coexist with high-capacity classical data traffic.

\begin{figure*}
    \centering
    \includegraphics[width=17.9 cm]{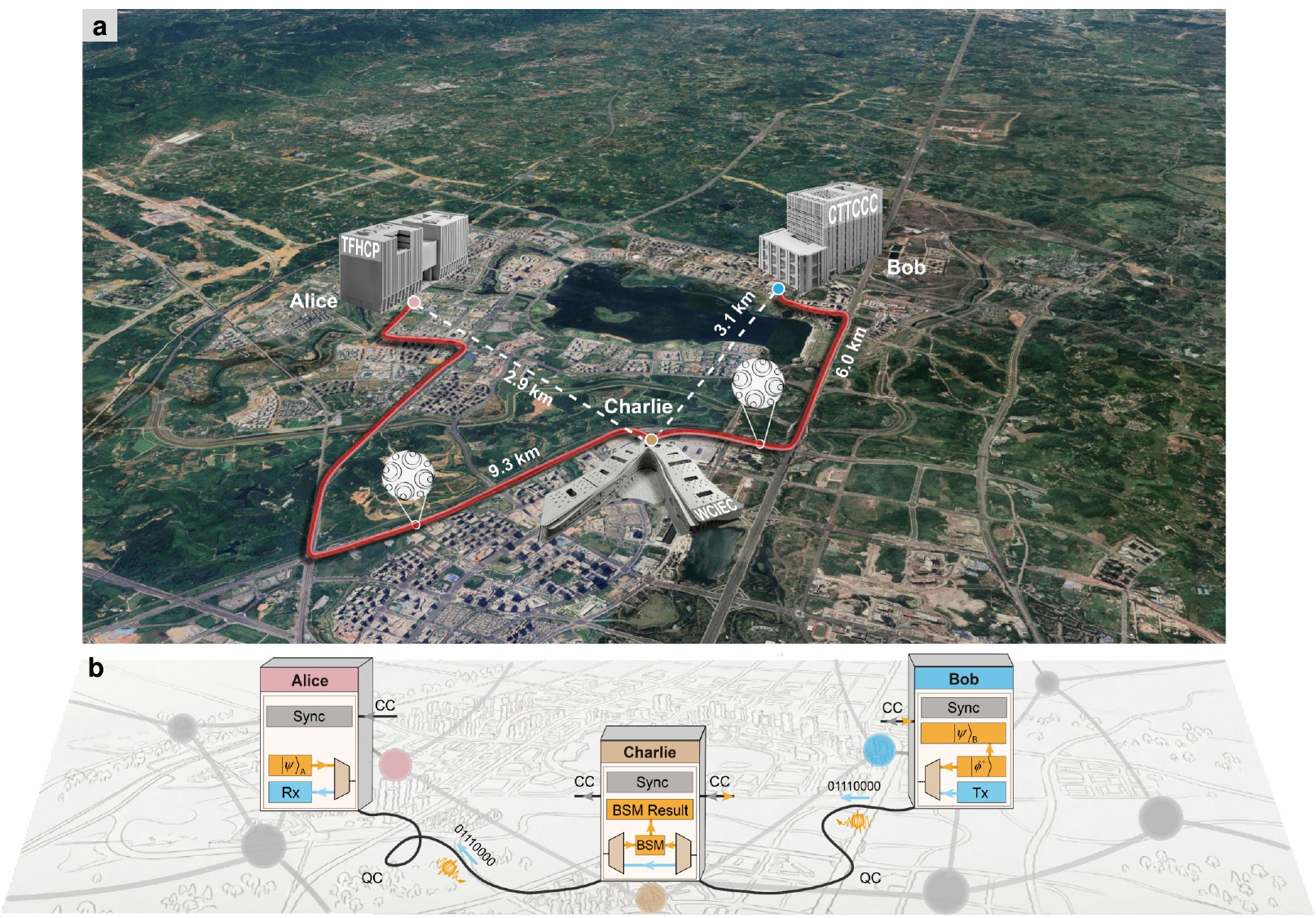}
    \caption{Deployed hollow core fibre (HCF) quantum teleportation network with classical data. \textbf{a}, Aerial view of the field-deployed metropolitan HCF network in Chengdu. Alice is located at Tianfu Haichuang Park (TFHCP), Bob at China Telecom Tianfu Cloud Computing Center (CTTCCC), and Charlie at Western China International Expo City (WCIEC). The three nodes are interconnected by deployed HCF links of 9.3~km between Alice and Charlie and 6.0~km between Bob and Charlie, respectively. 
    \textbf{b}, Conceptual diagram of the deployed HCF quantum teleportation system with co-propagating classical data. Alice prepares time-bin qubits $|\psi\rangle_A$ and sends them to Charlie through the HCF quantum channel (QC). Bob generates time-bin entangled photon pairs in $|\phi^+\rangle$, sends one photon through the HCF QC to Charlie for a Bell-state measurement (BSM), and analyses the other photon locally. A successful BSM teleports Alice’s input state onto Bob’s photon, up to the corresponding unitary correction. 
    Classical data are launched from Bob’s classical transmitter (Classical Tx), wavelength-multiplexed with the quantum signals into the deployed HCF QCs, demultiplexed and remultiplexed to bypass Charlie’s BSM setup, and finally received by Alice’s classical receiver (Classical Rx). 
    Synchronization signals and BSM results are transmitted through separate HCF classical channels (CCs) within the same deployed fibre infrastructure. 
Imagery \textcopyright 2026 Google. Map data \textcopyright 2026 Google.}
    \label{fig:Fig1}
\end{figure*}

Quantum teleportation coexisting with classical data traffic has only been demonstrated in SCF links by placing photonic qubits in the O-band, thereby reducing the spontaneous Raman scattering noise generated by high-power C-band telecom traffic \cite{thomas2024quantum}. This wavelength-engineering strategy enables teleportation in the presence of classical data traffic with launch powers up to 18.7 dBm. However, it merely mitigates rather than eliminates the nonlinear noise processes that originate from light-matter interactions in solid silica, leaving a fundamental constraint on quantum-classical coexistence in conventional fibre infrastructures. At the opposite extreme, to achieve the ultimate limit of low-interaction photonic transmission, a recent theoretical proposal explored lens-guided photonic transmission through evacuated tubes \cite{huang2024vacuum}. However, translating such vacuum beam guides from an idealized, physics-conceptual framework into scalable, field-deployable network infrastructure remains an elusive challenge, leaving a substantial gap between the theory and real-world utility.

Hollow core fibre (HCF), which guides light predominantly through air while preserving the practicality and scalability of fibre waveguide geometry \cite{petrovich2025broadband,shi2025all,sun2026polarization,carosini2026quantum}, offers a compelling platform for realizing low-interaction photonic transmission in deployed networks. Here, we demonstrate the first quantum teleportation over a field-deployed metropolitan HCF network situated around Xinglong Lake in Chengdu. Using telecom C-band photonic qubits, we achieve robust teleportation in the presence of co-propagating high-capacity classical data traffic, with classical launch powers of up to 160 mW. Our results establish air-guided HCF as a viable resource for shared quantum–classical communication infrastructures, combining low-noise quantum transmission with compatibility for conventional telecommunications. Beyond quantum teleportation, this capability provides a promising foundation for connecting heterogeneous quantum systems \cite{yu2020entanglement,van2022entangling,liu2026long,wang2026long} and for the deployment of repeater-enabled large-scale quantum networks.

\begin{figure*}
    \centering
    \includegraphics[width=18 cm]{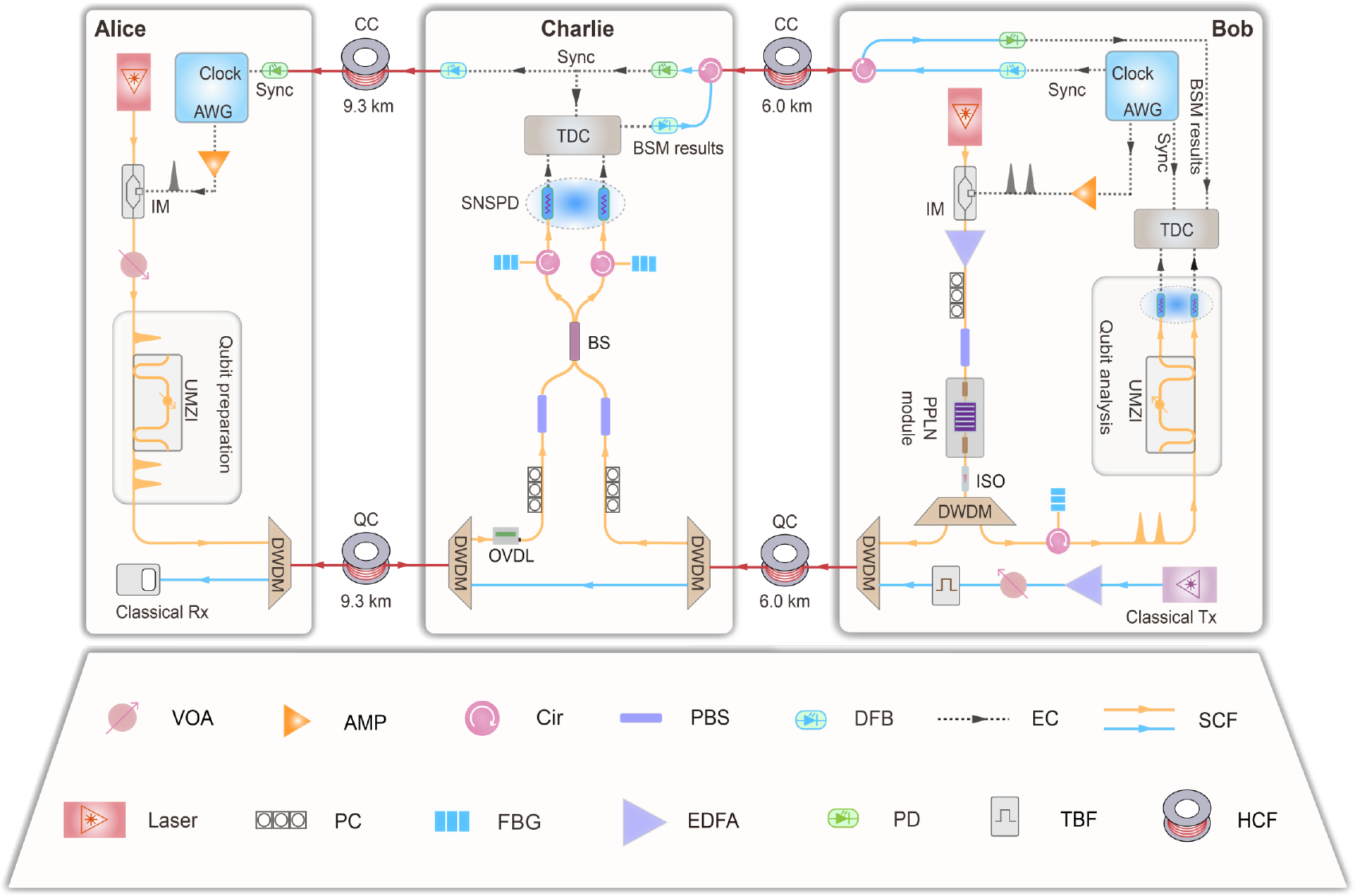}
    \caption{Schematics of experimental setup. 
    At Alice, laser pulses with a repetition rate of 500 MHz and a pulse width of 70 ps are sent through an unbalanced Mach-Zehnder interferometer (UMZI) with a path delay of 625 ps to prepare time-bin qubits in a superposition of early and late temporal modes. The drive signal of the intensity modulator (IM) is generated by an arbitrary waveform generator (AWG), amplified by an amplifier (AMP), and synchronized with Bob’s clock through CC. After modulation, the optical power is attenuated to the single-photon level by a digital variable optical attenuator (VOA). The prepared qubits are multiplexed into a 9.3‑km HCF QC to Charlie using a dense wavelength-division multiplexer (DWDM), with a total loss of 6.0 dB including contributions from fibre propagation, multiplexing, solid-core to hollow-core transitions, and hollow-core fusion splicing. See more details in Supplementary information S1.
    At Bob, time-bin entangled photon pairs are generated in a periodically poled lithium niobate (PPLN) waveguide module via cascaded second-harmonic generation (SHG) and spontaneous parametric downconversion (SPDC) processes. The pump laser is modulated by an IM into two pulses with a 625-ps separation and then amplified by an erbium-doped fibre amplifier (EDFA). A polarization controller (PC) and a polarization beam splitter (PBS) are used to ensure the polarization alignment for phase matching in the PPLN waveguide. The entangled photon pairs are separated by DWDM into signal and idler photons at 1532 nm and 1549 nm, respectively. The signal photons are filtered by a 10-GHz-bandwidth fibre Bragg grating (FBG) combined with an optical circulator (Cir) and then analysed with a UMZI providing a 625-ps delay matched to the time-bin separation. The idler photons are multiplexed and transmitted to Charlie through a 6.0-km-long HCF QC, with a total loss of 5.9 dB. 
    At Charlie, the idler photon interferes with the time-bin qubit from Alice at a 50:50 beam splitter (BS) for a Bell-state measurement (BSM). Two identical 10-GHz-bandwidth FBGs are applied before detection to spectrally select the interfering photons and suppress nonlinear noise. Within the deployed HCF QCs, quantum signals coexist with a classical communication channel at 1564.68 nm, both in the C-band, where the HCF introduces low nonlinear noise. 
    Classical data from Bob's Classical Tx are amplified by an EDFA, adjusted by a VOA, filtered through an 80‑GHz-bandwidth tunable bandpass filter (TBF), and then multiplexed with the idler photon via DWDM into the QC, with a launch power of 160 mW before multiplexing. The classical data is demultiplexed and remultiplexed to bypass Charlie's BSM setup, and finally received by Alice's Classical Rx, where the classical data is demultiplexed for monitoring with a power meter (PM), measuring approximately 10 mW after transmission through the deployed links. Synchronization (Sync) among all racks is referenced to the clock of the AWG at Bob and distributed via a separate HCF CC, which also carries the BSM results from Charlie to Bob. OVDL, optical variable delay line; BS, beam splitter; SNSPD, superconducting nanowire single photon detector; TDC, time-to-digital converter; ISO, isolator; DFB, distributed feedback laser; PD, photodetector; EC, electronic cables; SCF, solid core fibre.}
    \label{fig:Fig2}
\end{figure*}

 \begin{figure*}
    \centering
    \includegraphics[width=18 cm]{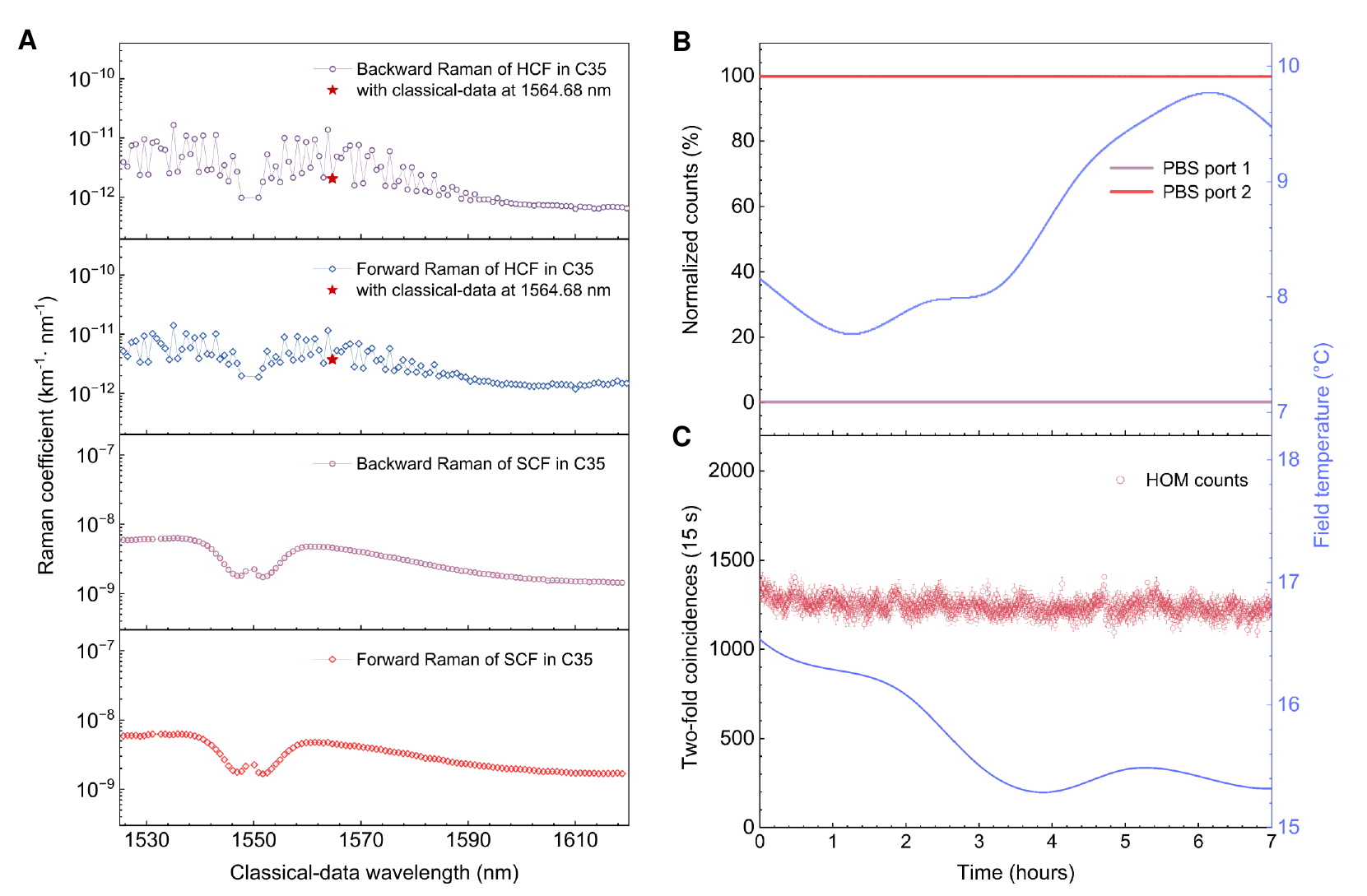}
    \caption{Characterization of the deployed HCF links. \textbf{a}, Raman scattering coefficients measured at the quantum channel wavelength of 1549 nm as the classical-data wavelength is scanned from 1525 to 1615~nm. 
    The upper two panels show the HCF results for backward scattering from the Alice-Charlie link and forward scattering from the Bob-Charlie link, which are the Raman noise components that enter the BSM and can affect the teleportation performance. The lower two panels show the corresponding measurements in SCF, which exhibit Raman scattering coefficients approximately three orders of magnitude higher than those of HCF. \textbf{b}, Passive polarization stability. A linearly polarized photon stream is sent through the fibre spool and analysed with a PBS after setting an initial polarization using a PC. After maximizing one PBS output (and thus minimizing the other), the ratio between the two PBS output ports varied by less than 2\% over seven hours. 
    \textbf{c}, Passive timing stability characterized by long-term Hong-Ou-Mandel (HOM) interference. The two-fold coincidence counts remain stable over seven hours, indicating preservation of the temporal overlap required for BSM. The right axis shows the temperature variation during the measurement. Error bars are calculated using Monte Carlo simulation, assuming Poissonian detection statistics. }
    \label{fig:Fig3}
\end{figure*}


\section{Deployed HCF infrastructure}
The aerial map of Chengdu, identifying the locations of Alice, Bob, and Charlie, is shown in Fig.~{\ref{fig:Fig1}}(a). Alice, the sender of the photonic qubits, located in Tianfu Haichuang Park (TFHCP), prepares attenuated laser pulses at a wavelength of 1549 nm with mean photon numbers $\mu_A \ll 1$ in various time-bin qubit states $|\psi\rangle_A = \alpha |e\rangle + \beta e^{i\phi} |l\rangle$, where $|e\rangle$ and $|l\rangle$ denote early and late temporal modes, respectively, $\phi$ is a phase factor, and $\alpha$ and $\beta$ are real numbers that satisfy $\alpha^2+\beta^2=1$. Using a 9.3-km HCF link, she sends her qubits to Charlie, who is located 2.9 km away in Western China International Expo City (WCIEC), as an intermediate node. Bob, the receiver of the qubits, located 3.1 km away in China Telecom Tianfu Cloud Computing Center (CTTCCC), himself creates pairs of idler and signal photons at 1549 nm and 1532 nm, respectively, in the maximally time-bin entangled state $|\phi^+\rangle = 2^{-1/2} \bigl(|e,e\rangle + |l,l\rangle\bigr)$. 

As shown in Fig.~{\ref{fig:Fig1}}(b) and Fig.~{\ref{fig:Fig2}}, the idler photons are multiplexed into a 6.0-km HCF link to co-propagate with classical data to the BSM node at Charlie, where they are probabilistically projected jointly with the photons from Alice onto the maximally entangled state $|\psi^-\rangle = 2^{-1/2} \bigl(|e,l\rangle - |l,e\rangle\bigr)$. The classical data is demultiplexed just before the BSM and then remultiplexed to bypass Charlie’s node and continue towards Alice, such that the BSM is performed under nonlinear-noise conditions generated through the deployed fibre links.

Once Bob receives the BSM results from Charlie, his signal photon acquires the state $|\psi\rangle_B = \sigma_y |\psi\rangle_A$, where $\sigma_y$ is the Pauli operator describing a bit-flip combined with a phase-flip. In other words, Charlie's measurement results in the teleportation of Alice's state, modulo a unitary transformation, through a three-node quantum network onto Bob's signal photon. Bob then performs a variety of projective measurements on this photon to confirm successful quantum teleportation, the outcomes of which, conditioned on a successful BSM at Charlie (Fig.~S9), are analysed using different approaches (see more details in Supplementary information S2). In this work, Bob’s signal photons are measured before the BSM, realizing a scenario in which teleportation is achieved a posteriori \cite{ma2012experimental,megidish2013entanglement,valivarthi2016quantum}.

\begin{figure*}
    \centering
    \includegraphics[width=18.2 cm]{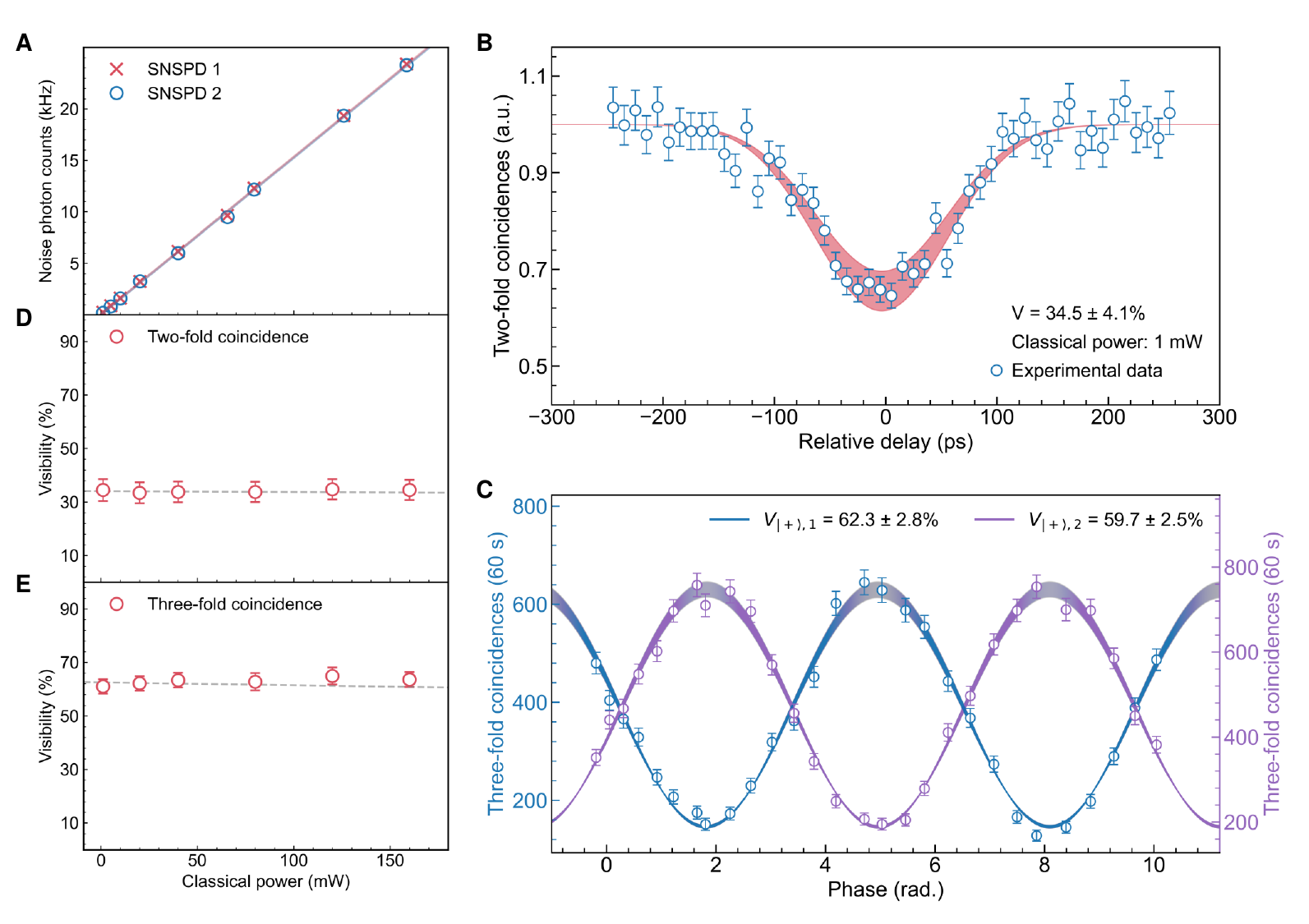}
    \caption{Quantum interference and teleportation under quantum-classical coexistence.
    \textbf{a}, Noise photon count rate measured by the two SNSPDs at Charlie as a function of the launched classical coexistence power.
    \textbf{b}, HOM interference measured at Charlie under a launched classical coexistence power of 1~mW. The fitted visibility is $34.5 \pm 4.1\%$. 
    \textbf{c}, Teleportation of a equatorial state under the same coexistence power. The purple and blue circles show the three-fold coincidence counts from the two output ports of Bob’s UMZI as a function of Bob’s analysis phase, with the average visibility of $61.0 \pm 2.7\%$. 
    \textbf{d}, Visibility of HOM interference as a function of the launched classical coexistence power. 
    \textbf{e}, Visibilities of sinusoidal three-fold coincidence counts as a function of the launched classical coexistence power measured for the $\lvert + \rangle$ state. 
 }
    \label{fig:Fig4}
\end{figure*}

We first evaluate the deployed HCF links as quantum channels by measuring nonlinear Raman noise, polarization fluctuations, and timing drifts under field conditions. By scanning the classical-data wavelength from 1525 to 1615~nm, we measure the induced noise photons falling within the C35 quantum channel, capturing forward scattering in the Bob-Charlie HCF link and backward scattering in the Alice-Charlie HCF link, respectively, both of which can enter the BSM and affect the teleportation performance.
As shown in Fig.~\ref{fig:Fig3}(a), the measured Raman scattering coefficient is approximately three orders of magnitude lower than that of SCF (See more details in Fig.~S5). This strong suppression of spontaneous Raman scattering reflects the reduced light-matter interaction in the air-guided core, which enables low-crosstalk coexistence between quantum signals and classical data channels within the same deployed fibre infrastructure.
The environmental stability of the deployed HCF links is further characterized through long-term polarization and quantum interference measurements. For polarization fluctuation, a continuous wave laser is transmitted through the HCF link and analysed using a polarization beam splitter (PBS) (Fig.~S3(a)). Figure~\ref{fig:Fig3}(b) shows the measured polarization extinction ratios of port 1 and port 2 at the 0:100 splitting ratio, which remains stable within 2\% over seven hours without active polarization compensation, despite continuous variations in the ambient temperature during the measurement. Comparable stability is observed for other polarization projection settings ranging from 10:90 to 50:50 (Figs.~S3(c-g)), indicating passive preservation of polarization states over extended field operation.
The timing drift is evaluated using long-term Hong-Ou-Mandel (HOM) interference measurements. As shown in Fig.~\ref{fig:Fig3}(c), the coincidence count rate remains stable over seven hours without active timing compensation under similar ambient temperature fluctuations. Consistent HOM interference traces recorded at one-hour intervals over a seven-hour period further confirm passive preservation of the temporal indistinguishability required for BSMs (Fig.~S4). 

\begin{figure*}
    \centering
    \includegraphics[width=18.2 cm]{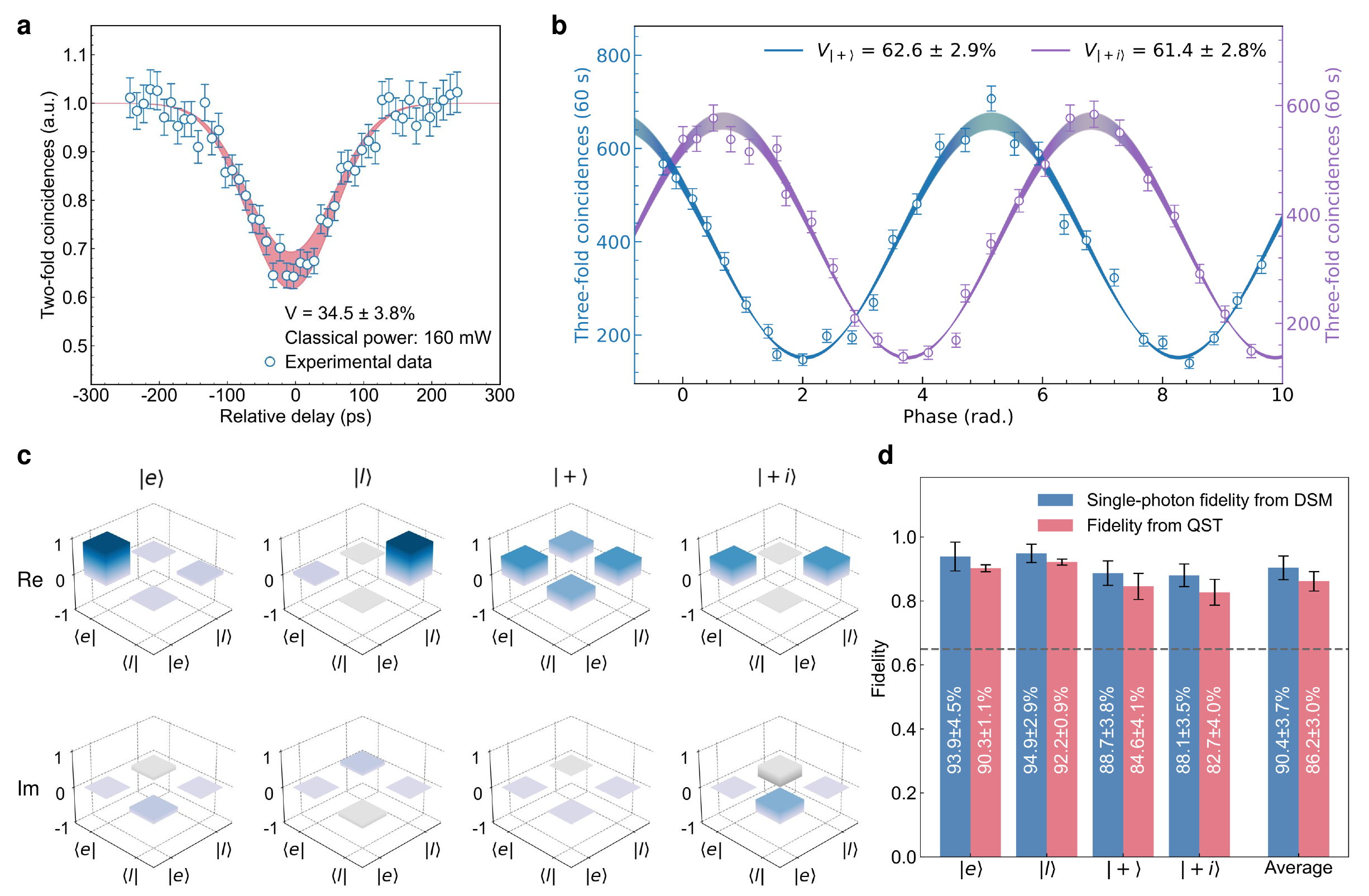}
    \caption{Quantum teleportation coexisting with classical launch power of 160 mW. 
    \textbf{a}, HOM interference with a visibility of $34.5 \pm 3.8\%$.
    \textbf{b}, Teleportation of the equatorial states. 
    The purple and blue circles represent the three-fold coincidence counts for states $|+\rangle$ and $|+i\rangle$, respectively, from a single output of the UMZI at Bob.     
    The average visibility of the fitting curves is $62.0 \pm 2.9\%$, indicating the coherence property of Alice’s state is successfully teleported to the signal photons. 
    \textbf{c}, Reconstructed density matrices of the teleported states $\lvert e\rangle$, $\lvert l\rangle$, $\lvert + \rangle$, and $\lvert +i \rangle$. The upper and lower panels show the real and imaginary parts, respectively. The state labels denote the states expected after teleportation.
    \textbf{d}, Teleportation fidelities for the four tested states and their average value. Purple bars show the fidelities obtained from quantum state tomography (QST), and blue bars show the corresponding single-photon fidelities extracted using the decoy-state method (DSM). In both cases, all fidelities exceed the classical limit of 2/3 (dashed line).  }
    \label{fig:Fig5}
\end{figure*}

\section{Quantum teleportation with data}
We next investigate whether the deployed HCF network preserves the indistinguishability of the photonic qubits and phase coherence required for quantum teleportation coexisting with classical data. As shown in Fig.~\ref{fig:Fig4}(a), the nonlinear Raman noise generated by the co-propagating classical channels is measured as a function of the launched classical coexistence power from 1 mW to 160 mW. 
The noise count rate increases weakly with the power, with a fitted slope of 153~Hz mW$^{-1}$, indicating strong suppression of nonlinear noise in the air-guided HCF links. Under a classical coexistence power of 1 mW, we verify the indistinguishability required for BSM through HOM interference measurements at Charlie. As shown in Fig.~\ref{fig:Fig4}(b), a visibility of $34.5 \pm 4.1\%$ is obtained, confirming that independently generated photons remain highly indistinguishable after transmission through the deployed HCF links. Quantum teleportation is then performed using time-bin qubits prepared by Alice in an equal superposition of $|e\rangle$ and $|l\rangle$ with a fixed phase, while Bob makes projection measurements onto states with various phases. Conditioned on a successful BSM at Charlie, the three-fold coincidence count rates exhibit clear sinusoidal interference as a function of Bob's phase as shown in Fig.~\ref{fig:Fig4}(c). The average three-fold coincidence rate reaches 10 Hz with a measured visibility of $61.0 \pm 2.7\%$, exceeding the maximum value achievable using the optimal classical strategy for single-photon teleportation \cite{massar1995optimal}. These results demonstrate the preservation of quantum coherence and successful quantum teleportation through the deployed HCF network under realistic quantum–classical coexistence conditions.
Figures~\ref{fig:Fig4}(d,e) show the extracted visibilities of HOM and phase coherence interference as functions of the launched classical power. Each data point is obtained by fitting the corresponding interference fringe measured under the specified coexistence condition. The HOM interference visibility remains close to 34\% across the entire power range (Tab.~S2), while the coherence visibility remains around 63\% (Tab.~S3). 

Finally, we test the teleportation performance under a classical coexistence power of 160~mW. Under this high-power condition, clear HOM interference and phase coherence interference fringes are still observed (Figs.~\ref{fig:Fig5}(a,b)), yielding visibilities of $34.5 \pm 3.8\%$ and $62.0 \pm 2.9\%$, respectively. These values remain consistent with those measured at lower coexistence powers, indicating that the deployed HCF links preserve the photonic qubits indistinguishability and phase coherence required for BSM and quantum teleportation in the presence of classical communication.
We then prepare and analyse photons in well-defined basis and superposition states, including $|e\rangle$, $|l\rangle$, $|+\rangle \equiv 2^{-1/2}(|e\rangle + |l\rangle)$ and $|+ i\rangle \equiv 2^{-1/2}(|e\rangle +i|l\rangle)$. This allows us to reconstruct the density matrices $\rho_{\mathrm{out}}$ after teleportation (Fig.~\ref{fig:Fig5}(c)), from which we calculate the fidelities $ F = {}_B\langle\psi|\rho_{\mathrm{out}}|\psi\rangle_B $ with the expected states $|\psi_B\rangle$. The fidelities for all four input states exceed the maximum classical value of 2/3 \cite{massar1995optimal}, with an average fidelity $\langle F \rangle = {(F_e + F_l + 2F_+ + 2F_{+i})}/{6} = 86.2 \pm 3.0\%$ (See more details in Tab.~S4). 
These results demonstrate high-fidelity quantum teleportation through the deployed HCF network while coexisting with high-power classical light.

The above fidelities certify the quantum nature of teleportation for attenuated coherent pulses, but the 2/3 bound is derived under a single-photon model. To extract the appropriate single-photon fidelity for the experiment, we therefore employ a decoy-state method (DSM) originally developed for quantum key distribution. Here, the DSM is used to characterize how the concatenation of the direct transmission from Alice to Charlie and the teleportation from Charlie to Bob, both implemented over HCF channels, impacts the teleportation fidelity of quantum states encoded into individual photons \cite{sinclair2014spectral}.
Alice varies her mean photon numbers of the signal, decoy, and vacuum states. For each setting, we independently perform quantum teleportation for $|e\rangle$, $|l\rangle$, $|+\rangle$, and $|+i\rangle$ (See more details in Tab.~S5).
The results (Fig.~\ref{fig:Fig5}(d)) show again that the fidelities for all tested states exceed the maximum value of 2/3 achievable in classical teleportation. 
Averaging over the tested states yields a single-photon teleportation fidelity of $\langle F^{(1)} \rangle \geq 90.4 \pm 3.7\%$, well above the classical limit. The deviations from unity mainly stem from residual photon distinguishability at the BSM and minor contributions from multi-photon emissions. These results confirm that field-deployed HCF links support high-fidelity quantum teleportation even under high-power quantum--classical coexistence conditions.

\section{Discussion and outlook}

Our results establish air-guided HCF as a promising channel platform for teleportation-based quantum networks. By guiding light predominantly within an air core while retaining the flexibility and deployability of fibre, HCF substantially eliminates the light–matter interactions that give rise to nonlinear noise and/or channel instability in conventional transmission media. Within our field-deployed metropolitan infrastructure, the HCF links maintain the indistinguishability of photonic qubits after transmission, enabling free-running BSM without active stabilization. These capabilities provide a robust, plug-and-play blueprint for scalable quantum internet \cite{valencia2026large}.
We also note that quantum teleportation in a field-deployed HCF network, where both quantum signals and classical data traffic operating in the C band, represents an important step from dedicated dark-fibre operation towards shared communication infrastructures. The air-guided architecture suppresses nonlinear noise photons by approximately three orders of magnitude relative to SCFs, enabling high-fidelity teleportation in the presence of classical launch powers up to 160 mW. Our scaling analysis further suggests that the coexistence regime could extend towards classical powers approaching 3 W (Fig.~S10(b)). These results highlight the potential of HCF as a common physical medium for high-capacity quantum–classical communication networks.

Looking forward, scaling quantum networks to continental distances fundamentally relies on quantum relays or repeaters to overcome exponential transmission losses. Although the loss of our current network is limited by first-generation deployed HCF links, recent demonstrations in ultra-low-loss HCF with attenuation below 0.1 dB km$^{-1}$ \cite{petrovich2025broadband,li2026low} suggest an over 100-km elementary link distance in future systems (Fig.~S11(b)). At the same time, the near-vacuum propagation velocity of HCF reduces transmission latency by approximately one-third relative to standard SCF, thereby relaxing coherence-time requirements for quantum memories \cite{wehner2018quantum,liu2021heralded,van2022entangling,wei2022towards,azuma2023quantum,liu2024creation,main2025distributed,delle2025operating,liu2026long}. Furthermore,  the 1532-nm photons employed in our experiment are naturally compatible, in both wavelength and bandwidth, with cryogenic erbium-based optical quantum memories \cite{erhan2014erQM,Zhang2023memory,jiang2023QM,wei2024memory,an2025quantum}, a key component of repeater-enabled quantum networks. Beyond this specific wavelength implementation here, air-guided fibre networks may provide a broader framework~\cite{zhang2025classical} for interconnecting heterogeneous quantum systems. Future HCF designs tailored to atomic, ionic, and solid-state optical transitions could reduce the reliance on efficiency-limited quantum frequency conversion interfaces~\cite{yu2020entanglement,van2022entangling,carosini2026quantum,liu2026long,wang2026long}. Such an approach may simplify the integration of disparate quantum technologies and facilitate the development of a large-scale heterogeneous quantum internet~\cite{kimble2008quantum,wehner2018quantum}.

\section{Methods}

\subsection{Fabrication and optical characterization of the HCF}

The HCF used in this work is an interstitial-tube-assisted double-nested antiresonant nodeless fibre (IT-DNANF), fabricated by the stack-and-draw process. A cross-sectional scanning electron microscope (SEM) image of the fabricated IT-DNANF is shown in Fig.~S1, with a core diameter of $28.1 \pm 0.7\,\mu\text{m}$. The average diameters of the nested tubes are $38.3 \pm 0.6\,\mu\text{m}$ for the large tubes, $28.5 \pm 0.8\,\mu\text{m}$ for the middle tubes and $11.5 \pm 0.6\,\mu\text{m}$ for the small tubes, while the interstitial tubes have an average diameter of $11.8 \pm 0.3\,\mu\text{m}$. The membrane thicknesses of the tubes are approximately $1160\,\text{nm}$, designed to centre the fundamental antiresonant transmission window near $1550\,\text{nm}$.
Within the C-band guidance window, the HCF exhibits broadband transmission with narrow absorption features associated with residual gas species in the air-filled core (Fig.~S2). 

\subsection{Entangled photon pairs}
Entangled photon pairs are generated in a fibre-pigtailed periodically poled lithium niobate (PPLN) waveguide module via cascaded second-harmonic generation (SHG) and spontaneous parametric downconversion (SPDC) processes \cite{zhang2021high}. The generated signal and idler photons are centred at 1531.90 nm and 1549.32 nm (rounded to 1532 nm and 1549 nm in the main text), corresponding to the centre wavelengths of ITU channels C57 and C35, respectively. Their properties are characterized by measuring single-photon and coincidence counts by varying pump power (Fig.~S7). In our experiment, the mean photon pair number is 0.031 per time bin with a 500-MHz repetition rate under a pump power of 6.2 mW. The coincidence-to-accidental ratio (CAR) is approximately 120. After transmission through the HCF links, the entanglement quality is verified by two-photon interference measurements under coexistence, as shown in Fig.~S8, and the visibilities are summarized in Tab.~S1.

\subsection{Indistinguishability without active compensation}
For a successful BSM, the photons arriving at Charlie from Alice and Bob must be indistinguishable in spatial, spectral, polarization, and temporal degrees of freedom. Spatial indistinguishability is ensured by the solid-core single-mode fibre components at the nodes. Spectral indistinguishability is achieved using two identical temperature-stabilized FBGs with a bandwidth of 10 GHz. Polarization overlap at Charlie is set using a pair of PBSs. Temporal overlap is calibrated via HOM interference, with the relative arrival time inferred from the coincidence dip due to the jitter of superconducting nanowire single photon detectors (SNSPDs) exceeding the wave-packet duration. The delay is coarsely adjusted via the electronic clock and finely tuned with an OVDL to minimize the dip, which serves as the zero-delay reference.

The maintenance of polarization and temporal overlap is ensured by the passive stability of HCF links. To assess the long-term polarization stability of the HCF links, a polarization-controlled photon stream is launched through the HCF and analysed with a PBS. The launched polarization is adjusted once to set the initial splitting ratio between the two PBS output ports and then left unchanged. The count ratio is monitored over extended operation periods, showing fluctuations below 2\% without active polarization stabilization (Fig.~S3). Based on this, long-term monitoring of the HOM coincidence counts confirmed stable arrival-time overlap without active timing compensation (Fig.~\ref{fig:Fig3}(c) and Fig.~S4).

\subsection{Quantum–classical coexistence architecture}
The quantum-classical coexistence link is implemented using cascaded spectral filtering, wavelength multiplexing, and demultiplexing. At Bob, the classical data is first filtered by a tunable bandpass filter (TBF) with an 80-GHz bandwidth and an isolation exceeding 50~dB, to suppress amplified spontaneous emission noise from the EDFA. The filtered classical data is then combined with Bob's idler photons using a DWDM centred at 1549.32~nm, with a 100-GHz bandwidth and an isolation of approximately 30~dB. The multiplexed idler photons and classical data are transmitted together through the 6.0-km HCF link from Bob to Charlie. At Charlie, the co-propagating signals are demultiplexed by a second DWDM with a 100-GHz bandwidth and an isolation of approximately 120~dB. The classical data is then remultiplexed by a third DWDM with a 100-GHz bandwidth and an isolation of 120 dB, into the 9.3-km HCF link, counter-propagating with the qubit prepared at Alice. At Alice, a fourth DWDM with a 100-GHz bandwidth and an isolation of approximately 30~dB demultiplexes the classical data from the quantum channel, after which the classical data is detected by the classical receiver (Classical Rx). 

For BSM, the idler photons demultiplexed from the second DWDM and the qubits demultiplexed from the third DWDM are further filtered by a 10-GHz-bandwidth FBG with an isolation of 30 dB before the SNSPDs. These FBGs are placed before the detectors at Charlie, rather than at Alice and Bob, providing improved tolerance to classical power. This cascaded filtering suppresses residual Raman noise, which is mainly contributed by forward Raman scattering from the 6.0-km Bob-Charlie HCF link and backward Raman scattering from the 9.3-km Alice-Charlie HCF link. 

\subsection{Synchronization} 

A common timing reference is required across the network to ensure that the generation, transmission, and detection of time-bin qubits remain aligned within the required timing precision. An electrical reference clock generated by an arbitrary waveform generator (AWG) is distributed to all nodes. Within each node, the electrical clock is routed through radio-frequency electronic cables (ECs) to synchronize the AWG and time-to-digital converter (TDC). For inter-node synchronization, the reference clock is converted into optical timing pulses using a distributed feedback (DFB) laser and transmitted through the CC HCF links. At each remote node, a photodetector (PD) converts the optical timing pulses back into an electrical clock, which is used to trigger the local AWG and TDC. Between Bob and Charlie, optical timing pulses and BSM result pulses travel in opposite directions through the same CC HCF link and are separated by optical circulators (Cir). This architecture establishes a shared temporal reference across the three-node HCF quantum network.

\subsection{Data collection}
Charlie performs a partial BSM by projecting Alice’s 1549-nm photon and the 1549-nm photon from Bob’s entangled pair onto $|\psi^-\rangle$. A successful projection is identified when the two SNSPDs at Charlie register photons in different time bins, with one detection in the early bin and the other in the late bin. The corresponding BSM heralding signals are transmitted to Bob through the classical channel and converted back into electrical pulses. At Bob, each BSM heralding signal is combined with the detection event of the 1532-nm photon after the projective analyser to form a three-fold coincidence. The relative delays among the heralding signals and Bob’s detection signals are calibrated and compensated using the programmable delay and coincidence modules of the TDCs. The TDCs continuously record the coincidence events, enabling automated acquisition of teleportation fringes, density-matrix measurements, and decoy-state data.
\\\\

\begin{acknowledgments}
\textbf{Acknowledgments}\\

This work was supported by Quantum Science and Technology-National Science and Technology Major Project (Nos.~2025ZD0300200, 2024ZD0300800), Sichuan Science Technology Program (Nos.~2025YFHZ0337, 2024YFHZ0368, 2024YFHZ0369, 2024YFHZ0370), the Frontier Technologies R\&D Program of Jiangsu (No.~BF2024036), National Natural Science Foundation of China (Nos.~62475039, 62405046), and Tianfu Jiangxi Laboratory (TFJX-ZD-2025-005).

We thank J. Zhang, J. Sun, Y. Wang, L.-P. Feng, and H.-M. Tan from China Telecom Co. Ltd., as well as H.-L. Cai from China Communications Services Co. Ltd., for their coordination and engineering support; Y. Qin and Y.-C. Shen from Zhongtian Technology Co. Ltd., as well as L.-M. Xiao from Fudan University, for HCF development and fabrication; Y.-G. Luo from China Telecom for live network integration; X. Ren from China Telecom for fibre cable deployment; T. Yang from China Communications Services Co. Ltd. for rack design; B. Lei and X.-L. Huo from China Telecom Co. Ltd. for constructive and insightful discussions; B. Xie from China Telecom Co. Ltd. and the supporting teams for implementation and operational support.
\\\\

\textbf{Data availability}\\

The datasets generated during this study are available from the corresponding author on reasonable request.
\\\\
\textbf{Author contributions}\\

The experiment was conceived by G.-C. G. and Q. Z. The experiment was guided by Y.-R. F., K. G., and Q. Z. The setup was developed, measurements were performed, and the data were analysed by R.-Y. S., Y.-Z. Z., Y.-R. F., Y.-B. M., S. S., Y.-Y. W., and Q. Z. The manuscript was written by Y.-R. F., R.-Y. S., Y.-Z. Z., and Q. Z. The SNSPDs were fabricated and tested by H. L. and L.-X. Y.



\textbf{Competing interests}\\

The authors declare no competing interests.
\\\\
\end{acknowledgments}


%

\end{document}